\documentclass[hyper,aps,amsmath,amssymb,nofootinbib]{revtex4}
\usepackage{lscape}

\usepackage{dsfont}
\newcommand{\bi}{\begin{itemize}}
\newcommand{\ei}{\end{itemize}}
\newcommand{\be}{\begin{eqnarray}}
\newcommand{\ee}{\end{eqnarray}}
\newcommand{\ba}{\begin{array}}
\newcommand{\ea}{\end{array}}
\newcommand{\bc}{\begin{center}}
\newcommand{\ec}{\end{center}}
\newcommand{\bt}{\begin{tabular}}
\newcommand{\btab}{\begin{table}}
\newcommand{\et}{\end{tabular}}

\begin{document}

\title{Transformations among large $c$ conformal field theories}

\author{Marcin Jankiewicz}
\email{m.jankiewicz@vanderbilt.edu}
\affiliation{Department of Physics and Astronomy, Vanderbilt
University, Nashville, TN 37235, USA}

\author{Thomas W. Kephart} 
\email{thomas.w.kephart@vanderbilt.edu}
\affiliation{Department of Physics and Astronomy, Vanderbilt
University, Nashville, TN 37235, USA}

\begin{abstract}\noindent
We show that there is a set of transformations that relates all of the 24 dimensional even self-dual (Niemeier) lattices, and also leads to non-lattice objects that however cannot be interpreted as a basis for the construction of holomorphic conformal field theory. In the second part of this paper, we extend our observations to higher dimensional conformal field theories build on extremal partition functions, where we generate $c=24k$ theories with spectra decomposable into the irreducible representations of the Fischer-Griess Monster. We observe interesting periodicities in the coefficients of extremal partition functions and characters of the extremal vertex operator algebras.
\end{abstract} 
\maketitle


\section{Introduction}
It was proven by Dixon et al. \cite{Dixon:1988qd} that the partition function $\mathcal{Z}$ of an arbitrary $c=24$ holomorphic conformal field theory based on $\mathds{R}^{24}/\Lambda$, where $\Lambda$ could be any of the 24 even self-dual Niemeier lattices in 24 dimensions, can be written as follows
\be\label{v-1}\mathcal{Z}=J+24(h+1)\,.\ee
Here $J$ is the weight zero modular function (with constant term set equal to zero), and $h$ is a Coxeter number for a lattice solution. We will show that any Niemeier lattice $\Lambda_{1}$, represented in terms of the $\Theta$-series related to the partition function (\ref{v-1}), can be obtained from another Niemeier $\Lambda_{2}$. We accomplish this using projections that rearrange points in the lattice to form a new lattice. Only for the particular combination of the projection parameters corresponding to the Coxeter numbers of the Niemeier lattices, do we have a lattice as a solution. For other combinations non-lattice solutions are obtained.\\

Any Niemeier lattice can be used as a starting point, i.e., any $\Theta$-function corresponding to a lattice can be used for the initial $\Theta$-function $Z_{0}^{0}$. The role of the transformation parameters is simple, they either rotate or rescale vectors in a lattice, moving some to different layers of the lattice. The number of transformation parameters depends strictly on an initial choice of the $\Theta$-function $Z_{0}^{0}$, and hence on the number of different conjugacy classes or the number of canonical sublattices in a lattice.
For example, for $E_{8}^{3}$, which is one of the Niemeier lattices, we have initially three parameters, one for each $SO(16)$ spinor conjugacy class. In the case of $D_{16}\times E_{8}$ initially we have $4$ parameters, one for the $SO(16)$ vector, spinor and conjugate spinor that build up the $D_{16}$ sublattice, and one parameter for the $SO(16)$ spinor of $E_{8}$. However, in these examples, upon constructing the new $\Theta$-series (or partition function), the initial number of parameters can be reduced to a single independent parameter, leading to the $\Theta$-function related to (\ref{v-1}) with $h$ being represented by the last free transformation parameter.\\

Parametrization of the twisted sector has been used to obtain new theories from 16 dimensional even self-dual lattices \cite{Blum:1997gw}. These include theories that were already known, like supersymmetric $E_{8}\times E_{8}$, nonsupersymmetric but non-tachyonic $SO(16)\times SO(16)$, and previously unknown theories like nonsupersymmetric and tachyonic $E_{8}\times SO(16)$, etc.\\

We generalize the analysis of \cite{Blum:1997gw}, to 24 dimensional lattices and also relax the constraints on the transformation parameters, i.e., we will no longer be working with $\mathds{Z}_{2}$ actions acting on the conjugacy classes, but rather with more complicated actions.\\
Since $\Theta$-functions of the 24 dimensional lattices are modular forms of weight $12$, and partition functions are modular invariant (weight zero), a natural question to ask is, which object is (more) physical? The answer to this question depends on the system being investigated.


\section{Modular Transformations and a 16 Dimensional Example}
We start with some basic definitions and a classic example already present in the literature \cite{Blum:1997gw}.
A formal definition of a $\Theta$-series of an even self-dual lattice $\Lambda$ is
\be\label{liar-1}\textsc{Z}_{\Lambda}=\sum_{x\in\Lambda}N(m)q^{m}\,,\ee
where $N(m)$ is a number of vectors with length squared equal to an even number $m$. From a mathematical point of view, $\textsc{Z}_{\Lambda}$ is a modular form of weight $\mathrm{dim}(\Lambda)/2$. Let us recall the difference between $\textsc{Z}_{\Lambda}$ and a partition function $\mathcal{Z}$.
Namely, a partition function $\mathcal{Z}$ is a modular function (a modular form of weight zero). By looking at the modular properties of $\textsc{Z}_{\Lambda}$ we conclude that it is related to $\mathcal{Z}$ by
$\mathcal{Z}=\textsc{Z}_{\Lambda}/\eta^{\mathrm{dim}(\Lambda)}\,,$
where
\be\eta(q)=q^{1/12}\prod_{m=1}^{\infty}(1-q^{2m})\ee
is a modular form of weight $1/2$ called Dedekind $\eta$-function with $q=e^{\pi i \tau}$ and $\tau$ is a modular parameter. For most of the time, we will focus on the lattices, hence we will work with $\Theta$-functions. However we will make some remarks about partition functions as well.\\

Let us investigate the relationship between $SO(32)$ and
$E_{8}\times E_{8}$ compactification lattices. The
$\Theta$-function of both can be expressed in terms of different
conjugacy classes of the $SO(16)\times SO(16)$ lattice, which is a
maximal common subgroup of both $SO(32)$ and $E_{8}\times E_{8}$.
$SO(2N)$ groups have four conjugacy classes namely, the $adjoint$
($I_{N}$), the $vector$ ($V_{N}$), the $spinor$ ($S_{N}$), and the
$conjugate~spinor$ ($C_{N}$). They can be expressed in terms of
Jacobi-$\theta$ functions as follows \cite{Lust:1989tj}:
\be\label{I}I_{N}\equiv\frac{1}{2}(\theta_{3}^{N}+\theta_{4}^{N})\,,\,\,V_{N}\equiv\frac{1}{2}(\theta_{3}^{N}-\theta_{4}^{N})\,,\ee
\be\label{S}S_{N}\equiv\frac{1}{2}\theta_{2}^{N}\,,\,\,C_{N}\equiv\frac{1}{2}\theta_{2}^{N}\,,\ee
where $N$ is the rank of $SO(2N)$. Both $spinor$ and $conjugate~spinor$ have the same $\Theta$-expansions.\\
Before going further, let us recall some of the properties of the modular group $\Gamma\simeq SL(2,\mathds{Z})$. It is generated by
\be\label{mod}T=\left(\begin{array}{cc}1&1\\0&1\end{array}\right)\,\,\mbox{and}\,\,S=\left(\begin{array}{cc}0&-1\\1&0\end{array}\right)\,,\ee
hence any element in this group can be written as $T^{n_{k}}ST^{n_{k-1}}S...ST^{n_{1}}$ where the $n_{i}$ are integers \cite{Apostol}.
The transformation rules for Jacobi-$\theta$ functions are given in (\ref{s2})-(\ref{t4}), and from them we obtain $\mathbf{S}$-transformed conjugacy classes
\be\label{mods-1}\textbf{S}(I_{N})=\frac{1}{2}(I_{N}+V_{N}+S_{N}+C_{N})\,,\,\,\textbf{S}(V_{N})=\frac{1}{2}(I_{N}+V_{N}-S_{N}-C_{N})\,,\ee
\be\label{mods-4}\textbf{S}(S_{N})=\frac{1}{2}(I_{N}-V_{N})\,,\,\,\textbf{S}(C_{N})=\frac{1}{2}(I_{N}-V_{N})\,,\ee
and $\mathbf{T}$-transformations acting on $\mathbf{S}$-transformed conjugacy classes of $SO(2N)$ give
\be\label{modt-1}\textbf{TS}(I_{N})=\frac{1}{2}(I_{N}-V_{N}+S_{N}+C_{N})\,,\,\,\textbf{TS}(V_{N})=\frac{1}{2}(I_{N}-V_{N}-S_{N}-C_{N})\,,\ee
\be\label{modt-4}\textbf{TS}(S_{N})=\frac{1}{2}(I_{N}+V_{N})\,,\,\,\textbf{TS}(C_{N})=\frac{1}{2}(I_{N}+V_{N})\,.\ee
We are going to use the $\textbf{TS}$ rather than the $\textbf{T}$ transformation in order to restore the modular invariance of the new partition functions resulting from the construction presented below.\\
Now consider the $\Theta$-function of the $SO(32)$ lattice, which
is given in terms of $SO(16)\times SO(16)$ conjugacy classes by
\be\label{part-so32}\textsc{Z}_{SO(32)}=(I_{8}^{2}+V_{8}^{2}+S_{8}^{2}+C_{8}^{2})\,,\ee
where squares are just a short-hand notation for $(I_{8},I_{8})$,
etc. A $\mathds{Z}_{2}$ action on the conjugacy classes means we
have to multiply a given conjugacy class by $\pm 1$. If one
chooses to act with transformations that flip the sign of the
vector and conjugate spinor representations of the second $SO(16)$
\cite{Blum:1997gw}, as shown in table \ref{tabel-1}
\bc\begin{table}[ht]\bc\begin{tabular}{|r|c|c|c|c|} \hline
                 & $I_{8}$ & $V_{8}$ & $S_{8}$ & $C_{8}$ \\ \hline
 First $SO(16)$  &  $+$   &   $+$  &    $+$ & $+$    \\
 Second $SO(16)$ &  $+$   &   $-$  &    $+$ & $-$    \\
 \hline
\end{tabular}\caption{\label{tabel-1}A $\mathds{Z}_{2}$ transformation applied to $\textsc{Z}_{SO(32)}$.}\ec
\end{table}\ec
one gets a new $\Theta$-series \be\label{part-tr}
\textsc{Z}_{1}=(I_{8}^{2}-V_{8}^{2}+S_{8}^{2}-C_{8}^{2})\,.\ee It
is obvious that $\textsc{Z}_{1}$ is not modular invariant as can
be seen from its $q$-expansion. I.e., some of the coefficients are
negative. In order to restore wanted modular properties, one has 
to add the $\textbf{S}$  and the $\textbf{TS}$ transformed forms
of $\textsc{Z}_{1}$:
\be\label{part-trs}\textbf{S}\textsc{Z}_{1}=I_{8}S_{8}+S_{8}I_{8}+C_{8}V_{8}+V_{8}C_{8}\,,\ee
\be\label{part-trst}\textbf{TS}\textsc{Z}_{1}=I_{8}S_{8}+S_{8}I_{8}-C_{8}V_{8}-V_{8}C_{8}\,.\ee
Adding partition functions (\ref{part-so32})-(\ref{part-trst}),
and also taking into account the overall normalization, one
obtains the $\Theta$-function of the dual $E_{8}\times E_{8}$
theory, namely
\be\label{part-e82}\frac{1}{2}\left(\textsc{Z}_{SO(32)}+\textsc{Z}_{1}+\textbf{S}\textsc{Z}_{1}+\textbf{TS}\textsc{Z}_{1}\right)=\textsc{Z}_{E_{8}\times
E_{8}}=(I_{8}+S_{8})^{2}\,.\ee


\section{Transformation Model in $24$ dimensions}
Now we move on to 24 dimensional lattices. We begin by writing the
$D_{16}\times E_{8}$ Niemeier lattice represented in terms of
$D_{8}\times D_{8}\times D_{8}$ conjugacy classes\footnote{We omit
subscript $``8"$ in the notation.}. We follow the standard
notation (see for example \cite{Dixon:1988qd}), where
$\textsc{Z}_{0}^{\,0}$ is the initial untwisted sector,
$\textsc{Z}_{0}^{\,1}$ is a projection,
$\textsc{Z}_{1}^{\,0}=\mathbf{S}\textsc{Z}_{0}^{\,1}$ and
$\textsc{Z}_{1}^{\,1}=\mathbf{T}\textsc{Z}_{1}^{\,0}$ are modular
transformed (twisted) sectors. Therefore
\be\label{w-1}\textsc{Z}_{0}^{\,0}=(I^{2}+V^{2}+S^{2}+C^{2})(I+S)\,,\ee
which explicitly, after evaluation of conjugacy classes in terms
of Jacobi-$\theta$ functions can be expanded into a $q$-series
\be\label{th-1}\textsc{Z}_{0}^{\,0}=1+720q^{2}+179280q^{4}+\mathcal{O}(q^{6})\,.\ee
Now let us make a projection, which in the most general way, can
be written as
\be\label{w-2}\textsc{Z}_{0}^{\,1}=(I^{2}+a~V^{2}+b~S^{2}+c~C^{2})(I+e~S)\,,\ee
where $a,b,c$ and $e$ are projection parameters, that change the
signs and/or scale the conjugacy classes. We also evaluate the
$q$-expansion of $\textsc{Z}_{0}^{\,1}$ to see the influence of
the projection we have just made
\be\label{th-2}\textsc{Z}_{0}^{\,1}=1+16(21+16a+8e)q^{2}+\mathcal{O}(q^{4})\,.\ee
This series is still even in powers of $q$, however self-duality
may have been lost. Using transformation properties
(\ref{mods-1})-(\ref{mods-4}) we get twisted sectors
\be\label{w-3}\textsc{Z}_{1}^{\,0}\equiv\mathbf{S}\textsc{Z}_{0}^{1}\!\!\!\!\!\!&=&\!\!\!\!\!\!\frac{1}{8}\left[(I+V+S+C)^{2}+a(I+V-S-C)^{2}+b(I-V)^{2}+c(I-V)^{2}\right]\times\cr&~~~~~~\times&\left[I+V+S+C+e~(I-V)\right]\,,\ee
and
\be\label{w-4}\textsc{Z}_{1}^{\,1}\equiv\mathbf{TS}\textsc{Z}_{0}^{1}\!\!\!\!\!\!&=&\!\!\!\!\!\!\frac{1}{8}\left[(I-V+S+C)^{2}+a(I-V-S-C)^{2}+b(I+V)^{2}+c(I+V)^{2}\right]\cr&~~~~~~\times&\left[I-V+S+C+e~(I+V)\right]\,.\ee
New transformed (orbifolded under certain circumstances) theories
can be represented in the following general form
\be\label{w-5}\textsc{Z}_{\mathrm{new}}=\frac{1}{2}\left(\textsc{Z}_{0}^{\,0}+\textsc{Z}_{0}^{\,1}+\textsc{Z}_{1}^{\,0}+\textsc{Z}_{1}^{\,1}\right)\,.\ee
We have chosen $\mathbf{S}$ and $\mathbf{TS}$ transformations to
restore modular invariance of a $\Theta$-function. Neither
$\textsc{Z}_{1}^{\,0}$ nor $\textsc{Z}_{1}^{\,1}$ are even. We can
see this directly from the $q$-expansion of
$\textsc{Z}_{1}^{\,0}$:
\be\label{z10}\textsc{Z}_{1}^{\,0}=\frac{1}{8}(1+a+b+c)(1+e)+2\left[2-b-c+e-3be-3ce+a(3+e)\right]q+\mathcal{O}(q^{2})\,.\ee
However, when added to $\textsc{Z}_{1}^{\,0}$, the $\mathbf{TS}$
transformed $\Theta$-function eliminates terms with odd powers in
$q$, i.e. its $q$-expansion is:
\be\label{z11}\textsc{Z}_{1}^{\,1}=\frac{1}{8}(1+a+b+c)(1+e)-2\left[2-b-c+e-3be-3ce+a(3+e)\right]q+\mathcal{O}(q^{2})\,.\ee
An even self-dual lattice scaled by the appropriate power of
Dedekind $\eta$-function is by definition a modular function, but
modular invariance does not necessarily imply we have a lattice;
therefore modular invariance is a more general concept.
This is important since only even lattices can be used as a basis for a compactification torus. But as we see we are safe; the odd terms in the twisted sectors have cancelled.\\
Note, $\textsc{Z}_{\mathrm{new}}$ is modular invariant regardless
of the values of the transformation parameters. However, a
$q$-expansion of a theory transformed in a way described above has
to be properly normalized. Normalization is not yet guaranteed,
since the zeroth order term in the $q$-expansion is given by
\be\label{start}1+\frac{1}{8}(1+a+b+c)(1+e)\,.\ee For $e=1$, we
are left with only one combination of the parameters $a,b$ and
$c$, such that normalization of the zeroth order term is fixed to
1 i.e., $a+b+c=-1$. As a result of this fixing, one is left with
\be\label{w-7}\textsc{Z}_{\mathrm{new}}=(I+S)^{3}\,.\ee The
resulting lattice corresponds to the $E_{8}\times E_{8}\times
E_{8}$ isospectral\footnote{A pair of lattices is said to be
isospectral if they have the same $\Theta$-expansion.} partner of
$D_{16}\times E_{8}$. But there is a more interesting case. For
$e=-1$ the normalization condition is already fixed, but this puts
no constraints on the rest of the parameters. Moreover, the
parameters $(a,b,c)$ are found in a specific
combination\footnote{For other values of $e$  the analysis gets
more complicated.} in every order in $q$
\be\label{w-8}x=3a-b-c\,.\ee We show this by explicitly evaluating
$\textsc{Z}_{\mathrm{new}}$
\be\label{zne}\textsc{Z}_{\mathrm{new}}=1+48(13+2x)q^{2}-144(-1261+16x)q^{4}+4032(4199+6x)q^{6}+...\,,\ee
which can be rewritten
\be\label{hmhm}\textsc{Z}_{\mathrm{new}}=\left[1+624q^{2}+181584q^{4}+16930368q^{6}+...\right]+96x\left[q^{2}-24q^{4}+252q^{6}+...\right]\,,\ee
where we have divided $\textsc{Z}_{\mathrm{new}}$ into $x$-dependent and $x$-independent parts. The number of independent parameters (after one fixes $e$) is reduced to one.
This parameter, $x$, can be related to the Coxeter number $h$ of a given lattice. The relation is model dependent and depends on an initial choice of $\textsc{Z}_{0}^{\,0}$. For the case at hand
\be\label{cox-1}x\equiv\frac{h-26}{4}\,.\ee
We observe that the $\Theta$-series in the first square bracket in (\ref{hmhm}) is an even self-dual (i.e. invariant under $\mathbf{S}$ and $\mathbf{T}$) function. It does not correspond to a lattice solution. However it can be related to the $J$-invariant (see (\ref{jay}) in Appendix-\ref{app-b}). Terms in the second square bracket in (\ref{hmhm}) form a unique cusp form of weight 12 \cite{Apostol}, which can be written as the $24$th power of the Dedekind $\eta$-function.
Using this knowledge we can write our solution in a more compact form
\be\label{result}\textsc{Z}_{\mathrm{new}}=\left[J+24(h+1)\right]\eta^{24}\,,\ee
where $h$ is a positive integer and is equal to the Coxeter number of the 24 dimensional even self-dual lattice if $\textsc{Z}_{\mathrm{new}}$ forms a Niemeier lattice.
Only for specific values of $x$ does one get a solution that corresponds to a lattice.
However, for the majority of cases representation in terms of group lattices is not possible. Physically this means that the gauge fields of the string theory (or CFT) do not transform under any gauge group, hence one is left with $24\cdot h$ singlets. Here we assume that the kissing number (or coefficient of the first nonvanishing $q$ term for non-lattice cases, which we will also call kissing number) is a multiple of $24$. \\

We now classify all solutions that can be derived by this technique. Using (\ref{cox-1}) in (\ref{result}) one finds the allowed values of $x$ form a set of $8191$ elements\footnote{For what it is worth $8191$ is a Marsenne prime.}. This is true under two conditions, first we assume that all the coefficients in a $q$-expansion are positive integers \cite{Conway}, \cite{Harvey:1988ur}. This assumption is not only reasonable but also physical, since these coefficients give us the number of states at each string mass level from the partition function point of view, and from the lattice point of view they correspond to the number of sites in each layer. The second assumption is that the ``kissing number'' for both lattices and non-lattices is an integer number which can change by one\footnote{One could assume that a kissing number for a non-lattice solution is still a multiple of $24$, as with lattices. Then modular properties of the $\Theta$-series are preserved, but the number of solutions is reduced to $342$.}.\\

Remembering that our starting point was the $D_{16}\times E_{8}$
lattice, one can immediately see that integer values of $(a,b,c)$
parameters correspond to ``relatives''\footnote{We call a pair of
group lattices relatives if they share a common maximal subgroup.}
of this lattice \cite{Conway}. Table \ref{t-2} shows the
transformation parameters for all five relatives of $D_{16}\times
E_{8}$. All except $E_{7}^{2}D_{10}$, can be gotten from the
action of a $\mathds{Z}_{2}$.
\begin{table}
\bc\begin{tabular}{|l|c|r|r|r|r|}
\hline
$Lattice$         & $h$   & $x$  & $a$  & $b$  & $c$  \\ \hline
$D_{24}$          & $46$  & $5$  & $1$  & $-1$ & $-1$ \\
$E_{8}^{3}$       & $30$  & $1$  & $1$  & $1$  & $1$  \\
$D_{12}^{2}$      & $22$  & $-1$ & $-1$ & $-1$ & $-1$  \\
$E_{7}^{2}D_{10}$ & $18$  & $-2$ & $0$  & $1$  & $1$ \\
$D_{8}^{3}$       & $14$  & $-3$ & $-1$ & $1$  & $-1$ \\
 \hline
\end{tabular}\ec
\caption{\label{t-2} Relatives of $D_{16}\times E_{8}$ lattice,
and their parametrization.}\end{table} Since $E_{8}^{3}$ and
$D_{16}\times E_{8}$ are isospectral, we can set $a=b=c=1$ in that
case. There are three other possible ``integer combinations'' for
$x$, corresponding to $D_{6}^{4}$, $D_{4}^{6}$ and $A_{1}^{24}$
(table \ref{t-3}), that are maximal subgroups of $D_{16}\times
E_{8}$.
\begin{table}[!h]
\bc\begin{tabular}{|l|c|r|r|r|r|}
\hline
$Lattice$            & $h$   & $x$     & $a$  & $b$  & $c$ \\ \hline
$D_{6}^{4}$          & $10$  & $-4$    & $-1$ & $1$  & $0$ \\
$D_{4}^{6}$          & $6$   & $-5$    & $-1$ & $1$  & $1$ \\
$A_{1}^{24}$         & $2$   & $-6$    & $-2$ & $1$  & $1$ \\
\hline
\end{tabular}\ec
\caption{\label{t-3} Maximal subgroups of $D_{16}\times E_{8}$
lattice, and their parametrization.}
\end{table}
Other integer values for $x$ that reach Niemeier lattices cannot be expressed in terms of integer values of $(a,b,c)$, which means that more complicated actions are needed. We list all the lattice solutions and the corresponding $x$ parametrization in table \ref{tab:finlis}. Other values of $(a,b,c)$ give the other Niemeier lattices and non-lattices solutions.\\
As a second example, we choose another Niemeier lattice
$E_{8}^{3}$ with a $\Theta$-function
\be\label{l-ex1}\textsc{Z}_{0}^{0}=(I+S)^{3}.\ee where the
projected sector is
\be\label{l-ex2}\textsc{Z}_{0}^{1}=(I+a~S)(I+b~S)(I+c~S)\,.\ee
Again to restore modular invariance we introduce twisted sectors
$\textsc{Z}_{1}^{0}$ and $\textsc{Z}_{1}^{1}$.
As a result a new modular invariant $\Theta$-function is obtained, and can be written in the form (\ref{result}), except that the relation between Coxeter number and transformation parameters is different. For example, after fixing $b=c=-1$ we are left with one free parameter $a=(h-22)/8$, which tells us that the space of the values of the $a$ parameters reaches all $8191$ solutions.\\
If one restricts parameters $(a,b,c)$ to $\pm 1$ one gets a family of simple $\mathds{Z}_{2}$ actions transforming the original lattice into the relatives  $D_{16}\times E_{8}$ and $D_{8}\times D_{8}\times D_{8}$ of $E_{8}^{3}$.\\

Let us finish this section with a following simple observation. The $\mathds{Z}_{2}$ orbifold actions are the actions which break/restore symmetry in a special way. If $\Lambda_{1}$ and $\Lambda_{2}$ have a common maximal subgroup then there is a $\mathds{Z}_{2}$ action that transforms $\Lambda_{1}$ into $\Lambda_{2}$. This is not the case when $\Lambda_{1}$ and $\Lambda_{2}$ do not have a common maximal subgroup. Therefore, this is possible for only a few pairs of Niemeier lattices (see table \ref{tab:fin}). This result follows the lines of the procedure discussed by Dolan et al. in \cite{Dolan:1989kf}.
\setcounter{table}{4}
\begin{table}[!h]\bc\begin{tabular}{ccccccccccc}
    $E_{8}D_{16}$ & $\rightarrow$ & $D_{8}^{3}$\\
               &             &  $\downarrow$ &             &            &             &            &             &       &    &   \\
    $E_{8}^{3}$  & $\rightarrow$ & $D_{8}^{3}$   & $\rightarrow$ & $D_{4}^{6}$  & $\rightarrow$ & $A_{1}^{24}$ & $\rightarrow$ & Leech          \\
    $D_{24}$     & $\rightarrow$ & $D_{12}^{2}$  & $\rightarrow$ & $D_{6}^{4}$  & $\rightarrow$ & $A_{3}^{8}$  &               & \\
    $E_{7}^{2}D_{10}$ & $\rightarrow$ & $A_{7}^{2}D_{5}^{2}$ & & & & & & \\
\end{tabular}\ec\caption{\label{tab:fin} Patterns of lattices obtained by $\mathds{Z}_{2}$ actions.}\end{table}

We want to emphasize that the main point of this section was to introduce transformations between existent lattices, and hence between known partition functions of their holomorphic conformal field theories. We do not claim that the other class of solutions we found (namely non-lattices) can be seen as new CFTs. They are however modular functions and can be seen as a left-over from the choice of the projection parameter which corresponds to the analog of a Coxeter number of the non-existent Niemeier lattice.
In the next section we discuss the generalization of these results to $k>1$. These examples will serve as a basis for construction of potential higher dimensional CFTs based on extremal $\Theta$-functions or partition functions.
The list of physical requirements which has to be imposed on a one-loop partition function of a conformal field theory was given for example in \cite{Harvey:1988ur} or in \cite{Dolan:1989vr}. All of these constraints are satisfied by any `candidate CFT' build on even self-dual lattice presented in this section and known earlier in the literature \cite{Dixon:1988qd}. Following this line of reasoning we proceed to discuss classes of CFTs build on extremal even self-dual lattices in dimensions $24k$ if they exist, or if not, on extremal even self-dual $\Theta$-series with $c=24k$. This avoids possible (or almost certain) complications of the construction based on non-lattice objects.


\section{CFT with $c=24k$}\label{app-a}
In this section we generalize our procedure to higher dimensions. We concentrate on lattices in $24k$ dimensions, since their $\Theta$-functions can be expressed in terms of positive integer\footnote{However, one can generalize this procedure to other dimensions ($8k$) as well.} powers of $\textsc{Z}_{\mathrm{new}}$ given in (\ref{result}). For example, one can use them in the construction of the lattices with dense packing in $48$ dimensions and the highest packing in $24$. These lattices are build on the so called extremal $\Theta$-functions.\\
For $k=1$ the kissing number (so the first non-zero coefficient in the $q$-expansion of the lattice) of the lattice with the highest packing is obtained as follows. The coefficients $a_{2}$ and $a_{4}$ in the $q$-expansion $1+a_{2} q^{2}+a_{4} q^{4}+...$ are constrained by the equation $24 a_{2}+a_{4}=196560$ (see below). From this we see that the maximum packing corresponds to the choice $a_{2}=0$ and $a_{4}=196560$.\\
In general, the equivalent of a 24 dimensional even self-dual, i.e. modular invariant lattice, can be obtained from
\be\eta^{24}(J+24+a_{2})=1+a_{2}q^{2}+(196560-24a_{2})q^{4}+252(66560+a_{2})q^{6}+...\ee
where $a_{2}$ is a positive integer. The extremal $a_{2}=0$ case is a Leech lattice. In order to preserve wanted properties we have to put constraints on values of the integer $a_{2}$. It is easy to see that $a_{2}\in\left[0,8190\right]$, generates $q$-expansion with positive entries.
The same kind of constraint can be imposed in 48 dimensions, where a modular invariant $\Theta$-series is written in the general form
\be\eta^{48}(J+24+a)(J+24+b)&=&1+(a+b)q^{2}+\left[2\cdot 196560-24(a+b)+ab\right]q^{4}+\cr&+&12\left[2795520+16401(a+b)-4ab\right]q^{6}+...\ee
By rewriting the expression as
\be1+a_{2}q^{2}+a_{4}q^{4}+(52416000+195660a_{2}-48a_{4})q^{6}+...\ee
we see that a dense packing with kissing number $52416000$ is obtained if one chooses $a_{2}=a_{4}=0$. It is the $P_{48}$ lattice \cite{Conway}. In 72 dimensions we have
\be1+a_{2}q^{2}+a_{4}q^{4}+a_{6}q^{6}+(6218175600+57091612a_{2}+ 195660a_{4}-72a_{6})q^{8}+...\,.\ee
In this case a lattice corresponding to the extremal $\Theta$-series would be obtained by setting $a_{2}$, $a_{4}$ and $a_{6}$ to zero so that the corresponding kissing number would be $6218175600$ except for the fact that this $\Theta$-series is not known to correspond to a lattice \cite{Sloane}.
The extremal $\Theta$-series in $24k$ dimensions obtained from this procedure is gotten by the requirement that all of the coefficients $a_{2},...,a_{2k}$ vanish\footnote{Alternatively, we can use the bound known in from the lattice theory \cite{Rains}, that the minimal norm of $n$-dimensional unimodular lattice is $\mu\leq \left[\frac{n}{24}\right]+2$.}. We can find in principle the number of solutions  with this parametrization, i.e., sensible $\Theta$-functions in $24k$ dimensions.
In $24$ dimensions the values of $a_{2}$ were constrained. In the rest of the cases, i.e., $k>1$, the number of independent parameters is $k$. However again the parameter space is finite. Using this information one can calculate the number of possible $\Theta$-functions in any dimension. For example in 48 dimensions we find $806022416786149$ $\Theta$-series. In this plethora of possibilities only some fraction can be interpreted as lattices, which can be related to CFT\footnote{For example there are $24^{2}$ even self-dual CFTs constructed out of 24 dimensional Niemeier lattices.}.  \\

Finally the partition function for any $24k$ dimensional theory contains a finite number of tachyons. For $k=1$ there is a single tachyon with $(m)^{2}=-1$, for $k>1$ we have $k$ tachyon levels in the spectrum. The most general formula of a partition function in $24k$ dimensions is
\be \mathcal{J}_{k}(\vec{x})\equiv\prod_{m=1}^{k}(J+24+x_{m})=\frac{1}{q^{2k}}\left[1+\sum_{m=(k-1)}^{\infty}f_{2m}(x_{1},...,x_{k})q^{2m-2k}\right]\,,\ee
where $\vec{x}=(x_{1},...,x_{k})$ and $f_{2m}\geq 0$ are polynomials in the $x_{i}$. The lowest (tachyonic) state with $(m)^{2}=-k$ is always populated by a single tachyon, and higher states are functions of $(x_{1},x_{2},...,x_{k})$. The $x_{i}$s can be chosen in such a way that all tachyon levels above the lowest level are absent, hence the next populated level would be occupied by massless states. This choice involves the elimination of $k-1$ parameters. The series would then depend on a single parameter $x_{k}$, more precisely the massless level is a polynomial in $x_{k}$ of the order $k$. The remainder of the spectrum does not depend on the choice of $x_{k}$, in analogy with the 24 dimensional case. What is appealing in these models is that for $k\gg 1$ we can have a single tachyonic state with arbitrarily large negative mass square that could potentially decouple from the spectrum leaving only states with $(m^{2})\geq0$, and the partition function of such a theory is still a well defined modular function. This may be an alternative to tachyon condensation \cite{Gerasimov:2000zp}. \\
Let us evaluate a few of examples with only a single tachyon coupled to the identity at $q^{2k}$ level for $k=1,2,3$ and $4$ which we define as $\mathcal{G}_{k}=\mathcal{J}_{k}|_{extremal}$. These are:
\be\label{tach}
\mathcal{G}_{1}(x_{1})\!&=&\!\frac{1}{q^{2}}+(24+x_{1})+196884q^{2}+...\cr
\mathcal{G}_{2}(x_{2})\!&=&\!\frac{1}{q^{4}}+(393192-48x_{2}-x_{2}^{2})+42987520q^{2}+...\cr
\mathcal{G}_{3}(x_{3})\!&=&\!\frac{1}{q^{6}}+(50319456-588924x_{3}+72x_{3}^{2}+x_{3}^{3})+2592899910q^{2}+...\cr
\mathcal{G}_{4}(x_{4})\!&=&\!\frac{1}{q^{8}}+(-75679531032-48228608x_{4}+784080x_{4}^{2}-96x_{4}^{3}-x_{4}^{4})+80983425024q^{2}+...\cr
&~&\ee
The allowed values of the polynomial coefficient of $q^{0}$ are integers that run from zero to the value of the $q^{2}$ coefficient.
If one changes $k$, then the $q^{2}$ coefficients are Fourier coefficients of the unique weight-2 normalized meromorphic modular form for $SL(2,\mathds{Z})$ with all poles at infinity \cite{tata}. The partition functions (\ref{tach}) are examples of the replication formulas \cite{Gannon:1999bi}.\\
There exists an interesting alternative set of CFTs with partition functions \cite{hohn} that we will call $\mathcal{H}_{k}$ where for different values of $k$ we can have
\be\label{char}
\mathcal{H}_{1}&=&\frac{1}{q^{2}}+196884q^{2}+...\cr
\mathcal{H}_{2}&=&\frac{1}{q^{4}}+1+42987520q^{2}+...\cr
\mathcal{H}_{3}&=&\frac{1}{q^{6}}+\frac{1}{q^{2}}+1+2593096794q^{2}+...\cr
\mathcal{H}_{4}&=&\frac{1}{q^{8}}+\frac{1}{q^{4}}+\frac{1}{q^{2}}+1+81026609428q^{2}+...\ee
Again we fix the tachyon levels by appropriate choices of the $x$s.
Note that the $q^{2}$ coefficients for $k=1$ and $2$ coincide in (\ref{tach}) and (\ref{char}) but not for larger $k$.
These are characters of the extremal vertex operator algebra of rank $24k$ (if it exists) \cite{hohn}. These characters were obtained by requiring $24k+\sum_{m=1}^{k}x_{m}=0$ (so that the $-(k-1)$ state is empty), all other coefficients of tachyon levels up the to massless states are fixed to one.\\

The extremal 24 dimensional case has been shown to be related to the Fischer-Griess monster group. In fact $\mathcal{G}_{1}(x_{1})$ is the modular function $j$ when $x_{1}=-24$. $j$ has the expansion
\begin{eqnarray}j=\frac{1}{q^{2}}+196884q^{2}+ 21493760q^{4} + 864299970q^{6} + 20245856256q^{8}+...\end{eqnarray}
\noindent
and the coefficients of this expansion decompose into dimensions of the irreducible representations of the monster (see table \ref{tab:mon2}),
\setcounter{table}{5}
\begin{table}
\begin{center}\begin{footnotesize}\begin{tabular}{|l|l|l|}
\hline
          & $J-invariant$      & $Monster$ \\ \hline
$j_{2}$   & $196884$           & $1+196883$    \\
          &                    &                \\
$j_{4}$   & $21493760$         & $1+196883+21296876$\\
          &                    &                    \\
$j_{6}$   & $864299970$        & $2\cdot 1+2\cdot 196883+21296876+842609326$\\
          &                    &                                            \\
$j_{8}$   & $20245856256$      & $3\cdot 1+3\cdot 196883+21296876+2\cdot842609326+18538750076$\\
          &                    &                                                              \\
$j_{10}$  & $333202640600$     & $4\cdot 1+5\cdot 196883+3\cdot 21296876+2\cdot 842609326+18538750076$\\
          &                    & $+19360062527+293553734298$\\
          &                    &\\
$j_{12}$  & $4252023300096$    & $3\cdot 1+7\cdot 196883+6\cdot 21296876+2\cdot 842609326+4\cdot 19360062527$\\
          &                    & $+293553734298+ 3879214937598$\\
\hline
\end{tabular}\end{footnotesize}\caption{\label{tab:mon2} Decomposition of the coefficients of $j$ into irreducible representations of the Monster group (for more see \cite{Borch}, \cite{He:2003pq}).}\end{center}
\end{table}
where we use the notation
\be j=\frac{1}{q^{2}}+j_{2}q^{2}+j_{4}q^{4}+...\ee
\begin{table}
\begin{center}\begin{footnotesize}\begin{tabular}{|c||l|l|l|l|l|l|l|l|l|l|l|}
\hline
$k$  & $g_{2}$   & $g_{4}$          & $g_{6}$         & $g_{8}$                & $g_{10}$        & $g_{12}$             \\ \hline
$2$  & $2j_{4}$  & $2j_{8}+j_{2}$   & $2j_{12}$       & $2j_{16}+j_{4}$        & $2j_{20}$       & $2j_{24}+j_{6}$      \\
$3$  & $3j_{6}$  & $3j_{12}$        & $3j_{18}+j_{2}$ & $3j_{24}$              & $3j_{30}$       & $3j_{36}+j_{4}$      \\
$4$  & $4j_{8}$  & $4j_{16}+2j_{4}$ & $4j_{24}$       & $4j_{32}+2j_{8}+j_{2}$ & $4j_{40}$       & $4j_{48}+2j_{12}$    \\
$5$  & $5j_{10}$ & $5j_{20}$        & $5j_{30}$       & $5j_{40}$              & $5j_{50}+j_{2}$ & $5j_{60}$            \\
$6$  & $6j_{12}$ & $6j_{24}+3j_{6}$ & $6j_{36}+2j_{4}$& $6j_{48}+3j_{12}$      & $6j_{60}$       & $6j_{72}+3j_{18}+2j_{8}+j_{2}$\\
$\vdots$ & $\vdots$        & $\vdots$                       & $\vdots$               & $\vdots$                & $\vdots$        & $\vdots$                         \\ \hline
$k$ & $g_{14}$   & $g_{16}$                  & $g_{18}$          & $g_{20}$           & $g_{22}$   & $g_{24}$             \\ \hline
$2$ & $2j_{28}$  & $2j_{32}+j_{8}$           & $2j_{36}$         & $2j_{40}+j_{10}$   & $2j_{44}$  & $2j_{48}+j_{12}$ \\
$3$ & $3j_{42}$  & $3j_{48}$                 & $3j_{54}+j_{6}$   & $3j_{60}$          & $3j_{66}$  & $3j_{72}+j_{8}$ \\
$4$ & $4j_{56}$  & $4j_{64}+2j_{16}+j_{4}$   & $4j_{72}$         & $4j_{80}+2j_{20}$  & $4j_{88}$  & $4j_{96}+2j_{24}+j_{6}$ \\
$5$ & $5j_{70}$  & $5j_{80}$                 & $5j_{90}$         & $5j_{100}+j_{4}$   & $5j_{110}$ & $5j_{120}$ \\
$6$ & $6j_{84}$  & $6j_{96}+3j_{24}$         & $6j_{108}+2j_{12}$& $6j_{120}+3j_{30}$ & $6j_{132}$ & $6j_{144}+3j_{36}+2j_{16}+j_{4}$ \\
$\vdots$ & $\vdots$        & $\vdots$                       & $\vdots$               & $\vdots$                & $\vdots$        & $\vdots$                 \\ \hline
$k$ & $g_{26}$   & $g_{28}$          & $g_{30}$           & $g_{32}$                & $g_{34}$       & $g_{36}$                  \\ \hline
$2$ & $2j_{52}$  & $2j_{56}+j_{14}$  & $2j_{60}$          & $2j_{64}+j_{16}$        & $2j_{68}$      & $2j_{72}+j_{18}$ \\
$3$ & $3j_{78}$  & $3j_{84}$         & $3j_{90}+j_{10}$   & $3j_{96}$               & $3j_{102}$     & $3j_{108}+j_{12}$ \\
$4$ & $4j_{104}$ & $4j_{112}+2j_{28}$& $4j_{120}$         & $4j_{128}+2j_{32}+j_{8}$& $4j_{136}$     & $4j_{144}+2j_{36}$ \\
$5$ & $5j_{130}$ & $5j_{140}$        & $5j_{150}+j_{6}$   & $5j_{160}$              & $5j_{170}$     & $5j_{180}$          \\
$6$ & $6j_{156}$ & $6j_{168}+3j_{42}$& $6j_{180}+2j_{20}$ & $6j_{192}+3j_{48}$      & $6j_{204}$     & $6j_{216}+3j_{54}+2j_{24}+j_{6}$\\
$\vdots$ & $\vdots$        & $\vdots$                       & $\vdots$               & $\vdots$                & $\vdots$        & $\vdots$            \\ \hline
$k$ & $g_{38}$   & $g_{40}$                  & $g_{42}$          & $g_{44}$           & $g_{46}$   & $g_{48}$                    \\ \hline
$2$ & $2j_{76}$  & $2j_{80}+j_{20}$          & $2j_{84}$         & $2j_{88}+j_{22}$   & $2j_{92}$  & $2g_{96}+j_{24}$            \\
$3$ & $3j_{114}$ & $3j_{120}$                & $3j_{126}+j_{14}$ & $3j_{132}$         & $3j_{138}$ & $3j_{144}+j_{16}$           \\
$4$ & $4j_{152}$ & $4j_{160}+2j_{40}+j_{10}$ & $4j_{168}$        & $4j_{176}+2j_{44}$ & $4j_{184}$ & $4j_{192}+2j_{48}+j_{12}$   \\
$5$ & $5j_{190}$ & $5j_{200}+j_{8}$          & $5j_{210}$        & $5j_{220}$         & $5j_{230}$ & $5j_{240}$                         \\
$6$ & $6j_{228}$ & $6j_{240}+3j_{60}$        & $6j_{252}+2j_{28}$& $6j_{264}+3j_{66}$ & $6j_{276}$ & $6j_{288}+3j_{72}+2j_{32}+j_{8}$     \\
$\vdots$ & $\vdots$        & $\vdots$                       & $\vdots$               & $\vdots$                & $\vdots$        & $\vdots$                         \\
 \hline
$k$ & $g_{50}$          & $g_{52}$                  & $g_{54}$          & $g_{56}$                  & $g_{58}$   & $g_{60}$ \\ \hline
$2$ & $2j_{100}$        & $2j_{104}+j_{26}$         & $2j_{108}$        & $2j_{112}+j_{28}$         & $2j_{116}$ & $2j_{120}+j_{30}$         \\
$3$ & $3j_{150}$        & $3j_{156}$                & $3j_{162}+j_{18}$ & $3j_{168}$                & $3j_{174}$ & $3j_{180}+j_{20}$         \\
$4$ & $4j_{200}$        & $4j_{208}+2j_{52}$        & $4j_{216}$        & $4j_{224}+2j_{56}+j_{14}$ & $4j_{232}$ & $4j_{240}+2j_{60}$         \\
$5$ & $5j_{250}+j_{10}$ & $5j_{260}$                & $5j_{270}$        & $5j_{280}$                & $5j_{290}$ & $5j_{300}+j_{12}$      \\
$6$ & $6j_{300}$        & $6j_{312}+3j_{78}$        & $6j_{324}+2j_{36}$& $6j_{336}+3j_{84}$        & $6j_{348}$ & $6j_{360}+3j_{90}+2j_{40}+j_{10}$      \\
$\vdots$ & $\vdots$               & $\vdots$                       & $\vdots$               & $\vdots$                       & $\vdots$      & $\vdots$      \\
 \hline
\end{tabular}\end{footnotesize}\caption{\label{tab-mon1} Coefficients of $24k$ dimensional extremal partition functions $\mathcal{G}_{k}$ in terms of coefficients $j_{2n}$ of modular function $j$.}\end{center}
\end{table}

For $24k$ one can expand the $q^{2n}$ coefficients of $\mathcal{G}_{k}$ in terms of $j$ coefficients which in turn can be expanded in terms of the dimensions of irreducible representations of the monster.
Table \ref{tab-mon1} demonstrates explicitly how the coefficients of the extremal $24k$ partition
functions are decomposed into the coefficients of
$j$.\\
Observe that the pattern of the $g_{2n}$ coefficients in the $k^{th}$ row in Table \ref{tab-mon1} is periodic with period $k$.
The first $k$ rows of the table of $g$ coefficients is overall $k!$ periodic.
We conjecture that this periodicity also continues to hold for all $k$.
The polynomial conditions to be satisfied to find the
extremal partition functions for large $k$ become increasingly more
difficult to solve with increasing $k$, so we do not have results
for $k>6$. Table \ref{tab-mon3} give the general periodicity.
\begin{table}
\begin{center}\begin{footnotesize}\begin{tabular}{|l|l||l|l|}
\hline
$k=2$        & $k=2$                                                 & $k=2$     & $k=2$     \\ \hline
$g_{4i+2}$   & $2j_{2(4i+2)}$                                        & $h_{4i+2}$& $2j_{2(4i+2)}$  \\
$g_{4i+4}$   & $2j_{2(4i+4)}+j_{2(2i+2)}$                            & $h_{4i+4}$& $2j_{2(4i+4)}+j_{2(2i+2)}$  \\ \hline

$k=3$        & $k=3$                                                 & $k=3$     & $k=3$  \\ \hline
$g_{6i+2}$   & $3j_{3(6i+2)}$                                        & $h_{6i+2}$& $3j_{3(6i+2)}+j_{6i+2}$  \\
$g_{6i+4}$   & $3j_{3(6i+4)}$                                        & $h_{6i+4}$& $3j_{3(6i+4)}+j_{6i+4}$  \\
$g_{6i+6}$   & $3j_{3(6i+6)}+j_{2i+2}$                               & $h_{6i+6}$& $3j_{3(6i+6)}+j_{2i+2}+j_{6i+6}$  \\ \hline

$k=4$        & $k=4$                                                 & $k=4$      & $k=4$  \\ \hline
$g_{8i+2}$   & $4j_{4(8i+2)}$                                          & $h_{8i+2}$ & $4j_{4(8i+2)}+2j_{2(8i+2)}+j_{8i+2}$  \\
$g_{8i+4}$   & $4j_{4(8i+4)}+2j_{2(2i+4)}$                           & $h_{8i+4}$ & $4j_{4(8i+4)}+2j_{2(2i+4)}+2j_{2(8i+4)}+j_{8i+4}+j_{4i+2}$  \\
$g_{8i+6}$   & $4j_{4(8i+6)}$                                        & $h_{8i+6}$ & $4j_{4(8i+6)}+2j_{2(8i+6)}+j_{8i+6}$  \\
$g_{8i+8}$   & $4j_{4(8i+8)}+2j_{(8i+8)}+j_{2i+2}$                   & $h_{8i+8}$ & $4j_{4(8i+8)}+2j_{(8i+8)}+j_{2i+2}+2j_{2(8i+8)}+j_{8i+8}+j_{4i+4}$  \\ \hline

$k=5$        & $k=5$                                                 & $k=5$       & $k=5$  \\ \hline
$g_{10i+2}$  & $5j_{5(10i+2)}$                                       & $h_{12i+2}$ & $g_{12i+2}+3j_{3(12i+2)}+2j_{2(12i+2)}+j_{12i+2}$    \\
$g_{10i+4}$  & $5j_{5(10i+4)}$                                       & $h_{12i+4}$ & $g_{12i+4}+3j_{3(12i+4)}+2j_{2(12i+4)}+j_{12i+4}+j_{6i+2}$      \\
$g_{10i+6}$  & $5j_{5(10i+6)}$                                       & $h_{12i+6}$ & $g_{12i+6}+3j_{3(12i+6)}+2j_{2(12i+6)}+j_{12i+6}+j_{4i+2}$   \\
$g_{10i+8}$  & $5j_{5(10i+8)}$                                       & $h_{12i+8}$ & $g_{12i+8}+3j_{3(12i+8)}+2j_{2(12i+8)}+j_{12i+8}+j_{6i+4}$    \\
$g_{10i+10}$ & $5j_{5(10i+10)}+j_{2i+2}$                             & $h_{12i+10}$& $g_{12i+10}+3j_{3(12i+10)}+2j_{2(12i+10)}+j_{12i+10}$  \\
             &                                                       & $h_{12i+12}$& $g_{10i+12}+3j_{3(12i+12)}+2j_{2(12i+12)}+j_{12i+12}+j_{6i+6}+j_{4i+4}$\\ \hline
$k=6$        & $k=6$                                                 &   & \\ \hline
$g_{12i+2}$  & $6j_{6(12i+2)}$                                       &  &  \\
$g_{12i+4}$  & $6j_{6(12i+4)}+3j_{3(6i+2)}$                          &  &  \\
$g_{12i+6}$  & $6j_{6(12i+6)}+2j_{2(4i+2)}$                          &  &  \\
$g_{12i+8}$  & $6j_{6(12i+8)}+3j_{3(6i+4)}$                          &  & \\
$g_{12i+10}$ & $6j_{6(12i+10)}$                                      &  & \\
$g_{12i+12}$ & $6j_{6(12i+12)}+3j_{3(6i+6)}$                         &  & \\
             & $+2j_{2(4i+4)}+j_{2i+2}$                              &  & \\ \hline
 \end{tabular}\end{footnotesize}\caption{\label{tab-mon3} Periodicity of the coefficients $g_{n}$ for $c=24k$ extremal partition functions $\mathcal{G}_{k}$, and for $h_{n}$ coefficients of characters of the extremal vertex operator algebras $\mathcal{H}_{k}$ in terms of coefficients the $j_{2n}$ of the modular function $j$  ($k=6$ case for $h_{n}$ is not displayed since it is long but it has period 24).}\end{center}\end{table}

These results are somewhat reminiscent of Bott periodicity for the
stable homotopy of the classical groups. Here we are dealing with (the
equivalent of) increasing level algebras.

 To summarize, when $k=1$ it is known via standard Monster
Moonshine that the coefficients of $j$
decompose into Monster representations \cite{Gannon:1999bi}. The related extremal lattices are Leech in $24$ and $P_{48}$ in $48$ dimensions.  The fact that all the higher $k$
coefficients also decompose into Monster representations indicates
that they have large symmetries containing the Monster and
the fact that they have these symmetries may indicate that they are related to $24k$ dimensional lattices.


\section{Open Questions/Problems}

In the discussion section of \cite{Dixon:1988qd} it is argued that by using $\mathds{Z}_{2}$ twists of the 23 Lie type Niemeier lattices one gets holomorphic conformal field theories that are not graded isomorphic to any of the untwisted theories based on Niemeier lattices. What we have shown is that there exists a family of transformations, not necessarily of the simple $\mathds{Z}_{2}$ form, that connects the class of holomorphic conformal field theories the Niemeier lattices. Furthermore these transformations connect non-Niemeier even self-dual $c=24$ $\Theta$-functions. These results generalize to any $c=24k$ case (in particular, when the resulting parametrization corresponds to $24k$ dimensional lattice). We also found interesting patterns of periodicity related to the Monster moonshine.\\
There exists a possible application of conformal field theories (with high central charge) to cosmology, since for $k\rightarrow\infty$, $(m_{tachyon})^{2}\rightarrow-k
\cdot M_{Planck}^{2}\rightarrow-\infty$, which suggests this tachyon maybe the single tachyon of $\mathcal{J}(x_{k})$ and lead to some variant of a tachyon condensation \cite{Gerasimov:2000zp}.
Also in the case of $k\rightarrow\infty$ (hence divergent central charge \cite{Bilal:1988jf}, \cite{Verlinde:2000wg}) we get, for example, a theory with a gauge lattice $G^{k}$. Depending on the representation content of the gauge group one could potentially partially deconstruct $G^{k}$ \cite{Arkani-Hamed:2001ca} to go from $2D$ CFT to a $4D$ theory.\\
It is important to mention that some of the results obtained here (related to the transformations between $c=24$ CFTs) were partially present in the literature \cite{Dolan:1989kf}, \cite{Dolan:1994st}. For example the $\mathds{Z}_{2}$ patterns (in table \ref{tab:fin}) were explained in \cite{Dolan:1994st}. Also it was shown that from any even-self dual lattice it is always possible to construct one untwisted and twisted conformal field theory.   However in our case it is enough to start from just a single even self-dual lattice to obtain all other lattice solutions by a proper choice of projection parameters.
Finally there are a number of interesting open questions. For instance, is there any relation between solutions found here (including non-lattice solutions) and other solutions corresponding to higher level Kac-Moody algebras classified by Schellekens in \cite{Schellekens:1992db}, and others \cite{Montague:1994yb} and \cite{Montague:1997nu}? Is there an interpretation of non-lattice solutions? Can one interpret them as lattices on spaces with curvature?\\ The natural place for theories with $c>24$ is in condensed matter systems, since there are not enough ghosts to cancel a conformal anomaly. Nevertheless, we hope applications of the transformation techniques investigated here may lead to further/deeper understanding of dualities relating $\mathcal{N}=2$ heterotic string theories in $2D$ \cite{Vafa2}.
\section*{Acknowledgments}
We benefited greatly from several discussions with John Ratcliffe. We also thank Rich Holman, Burt Ovrut, Ashoke Sen and Hirosi Ooguri for discussions, Neil Sloane for an email correspondence, and Ralf Lehnert for careful reading of the manuscript.
This work is supported in part by U.S. DoE grant \#~DE-FG05-85ER40226.

\appendix

\section{Relations between modular functions and lattices}\label{app-b}
The $\textsc{Z}_{\Lambda}$ of any lattice $\Lambda$ can be expressed in terms of the Jacobi $\theta$-functions:
$$\theta_{3}(\xi|\tau)\equiv \sum_{m=-\infty}^{\infty}e^{2mi\xi+\pi i \tau m^{2}}\,.$$
For us it will be enough to work with the simpler theta functions, often called $\theta$ constants, defined as:
\be&~&\theta_{2}(\tau)\equiv e^{\pi i \tau/4}\theta_{3}(\frac{\pi\tau}{2}|\tau)\,,\,\,\theta_{3}(\tau)\equiv\theta_{3}(0|\tau)\,,\nonumber\\
&~&\theta_{4}(\tau)\equiv\theta_{3}(\tau+1)\,.\ee
These functions have fantastic properties including a basically infinite web of identities which will be used later on. Most important for us, is that they have very simple modular transformation properties.  Here we show their behavior under the generators of the modular group, namely under $\mathbf{S}$ we have
\be\label{s2}\theta_{2}(-1/{\tau})=\left(\frac{\tau}{i}\right)^{\frac{1}{2}}\theta_{4}(\tau)\,,\ee
\be\label{s3}\theta_{3}(-1/{\tau})=\left(\frac{\tau}{i}\right)^{\frac{1}{2}}\theta_{3}(\tau)\,,\ee
\be\label{s4}\theta_{4}(-1/{\tau})=\left(\frac{\tau}{i}\right)^{\frac{1}{2}}\theta_{2}(\tau)\,,\ee
and under $\mathbf{T}$:
\be\label{t2}\theta_{2}(\tau+1)=\sqrt{i}\theta_{2}(\tau)\,,\ee
\be\label{t3}\theta_{3}(\tau+1)=\theta_{4}(\tau)\,,\ee
\be\label{t4}\theta_{4}(\tau+1)=\theta_{3}(\tau)\,.\ee
These transformation rules are easily derived using the Poisson resummation formula \cite{Conway}.
Finally, we introduce the modular invariant function $J$, which plays a very important role in our considerations
\be\label{jay}J&\equiv&\frac{1}{\eta^{24}}\left[(\theta_{3}(\tau)\theta_{4}(\tau))^{12}+(\theta_{2}(\tau)\theta_{3}(\tau))^{12}-(\theta_{2}(\tau)\theta_{4}(\tau))^{12}\right]+24\cr
&=&\frac{1}{q^{2}}+196884q^{2}+21493760q^{4}+\mathcal{O}(q^{6})+...\,.\ee
This function (sometimes called the $J$-invariant related to weight-zero modular function $j$ by $J=j-744$) is the modular form of weight zero, as can be easily proved using transformations (\ref{s2})-(\ref{t4}). In the denominator of (\ref{jay}) we have the $24$th power of Dedekind $\eta$-function
\be\label{change-3}\eta^{24}(\tau)=\left[q^{1/12}\prod_{m=1}^{\infty}(1-q^{2m})\right]^{24}=q^{2}-24q^{4}+252q^{6}-1472q^{8}+...\,,\ee
which is the unique form of weight $12$, with the following transformation rules under $\mathbf{S}$ and $\mathbf{T}$
\be\label{tr-eta}\eta^{24}(-1/{\tau})=(\sqrt{-i\tau}\eta(\tau))^{24}\,,\,\,\eta^{24}(\tau+1)=(e^{i\pi/12}\eta(\tau))^{24}\,.\ee




\begin{turnpage}
\setcounter{table}{3}
\begin{table}\small\begin{tabular}{|l||l|l|l|r|r|r|}
  \hline
Lattice              & Coxeter & $x$  & Glue Code                                                               & $a_{2}$  & $a_{4}$& $a_{6}$ \\ \hline
$E_{8}$              & $30$    &      & $+$                                                                     & $240$ & $2160$   & $6720$       \\
\hline
$E_{8}^{2}$          & $30$    &      & $++$                                                                    & $480$ & $61920$  & $1050240$    \\
$D_{16}$             & $30$    &      & $+$                                                                     & $480$ & $61920$  & $1050240$    \\
\hline
$D_{24}$             & $46$    & $5$  & $+$                                                                     & $1104$ & $170064$ & $17051328$  \\
$D_{16}E_{8}$        & $30$    & $1$  & $++$                                                                    & $720$  & $179280$ & $16954560$  \\
$E_{8}^{3}$          & $30$    & $1$  & $+++$                                                                   & $720$  & $179280$ & $16954560$  \\
$A_{24}$             & $25$    &$-1/4$& $[0]+2([5]+[10])$                                                       & $600$  & $182160$ & $16924320$  \\
$D_{12}^{2}$         & $22$    &$-1$  & $[00]+2[12]+[11]$                                                       & $528$  & $183888$ & $16906176$  \\
$A_{17}E_{7}$        & $18$    &  $-2$& $[00]+2[31]+2[60]+[91]$                                                 & $432$  & $186192$ & $16881984$  \\
$D_{10}E_{7}^{2}$    & $18$    & $-2$ & $[000]+[110]+[310]+[211]$                                               & $432$  & $186192$ & $16881984$  \\
$A_{15}D_{9}$        & $16$    &$-5/2$& $[00]+2([12]+[16]+[24])+[08]$                                           & $384$  & $187344$ & $16869888$  \\
$D_{8}^{3}$          & $14$    &$-3$  & $[000]+3([011]+[122])+[111]$                                            & $336$  & $188496$ & $16857792$  \\
$A_{12}^{2}$         & $13$    &$-13/4$& $[00]+4([15]+[23]+[46])$                                                & $312$  & $189072$ & $16851744$  \\
$A_{11}D_{7}E_{6}$   & $12$    &$-7/2$ & $[000]+2[111]+2[310]+2[401]+2[511]+[620]$                               & $288$  & $189648$ & $16845696$  \\
$E_{6}^{4}$          &  $12$   &$-7/2$ & $[0000]+8[1110]$                                                        & $288$  & $189648$ & $16845696$  \\
$A_{9}^{2}D_{6}$     &  $10$   &$-4$ & $[000]+4([121]+[132]+[240]+[341])+2[051]+[552]$                         & $240$  & $190800$ & $16833600$  \\
$D_{6}^{4}$          &  $10$   &$-4$ & $[0000]+12[0123]+[1111]+[2222]+[3333]$                                  & $240$  & $190800$ & $16833600$  \\
$A_{8}^{3}$          &  $9$    &$-17/4$& $[000]+6([114]+[330]+[122]+[244])+2[333]$                               & $216$  & $191376$ & $16827552$  \\
$A_{7}^{2}D_{5}^{2}$ &  $8$    &$-9/2$ & $[0000]+4([0211]+[1112]+[2202]+[3312]+[2411])+8[1301]+2[0422]+[4400]$   & $192$  & $191952$ & $16821504$  \\
$A_{6}^{4}$          &  $7$    &$-19/4$& $[0000]+24[0123]+8([1112]+[1222]+[2223])$                               & $168$  & $192528$ & $16815456$  \\
$A_{5}^{4}D_{4}$     &  $6$    &$-5$ & $[00000]+6[00331]+24([00121]+[12231])+8([02220]+[11130])+[33330]$       & $144$  & $193104$ & $16809408$  \\
$D_{4}^{6}$          &  $6$    &$-5$ & $[000000]+45[000011]+18[111111]$                                          & $144$  & $193104$ & $16809408$  \\
$A_{4}^{6}$          &  $5$    &$-21/4$& $[000000]+60[001122]+40[111222]+12[011111]+12[022222]$                  & $120$  & $193680$ & $16803360$  \\
$A_{3}^{8}$          &  $4$    &$-11/2$& $[3(2001011)]+cyclic~permutations~of~(2001011)$                                                          & $96$   & $194256$ & $16797312$  \\
$A_{2}^{12}$         &  $3$    &$-23/4$& $[2(11211122212)]+cyclic~permutations~of~(11211122212)$                                                      & $72$   & $194832$ & $16791264$  \\
$A_{1}^{24}$         &  $2$    &$-6$  & $[1(00000101001100110101111)]+cyclic~permutations~of~(00000101001100110101111)$                                           & $48$   & $195408$ & $16785216$  \\
$Leech$              &  $0$    &$-13/2$& $None$                                                                  & $0$    & $196560$ & $16773120$  \\

  \hline

\end{tabular}

\caption{\label{tab:finlis} All lattice solutions in 24 dimensions. The second column represents the Coxeter number $h$. The third column shows $x$ parametrization when $Z_{0}^{0}=Z_{D_{16}E_{8}}$. The glue code in the explicit form is given for most of the lattices and in a generator form for $A_{3}^{8}$, $A_{2}^{12}$, $A_{1}^{24}$ as in \cite{Conway}. Numbers in this column represent conjugacy classes, $+$ means combination of adjoint and spinor conjugacy classes, $0$ stands for the adjoint, $1,2,3$ are vector, spinor and conjugate spinor for $D_{n}$ lattices (similarly for $A_{n}$ with $n-1$, and $E_{6}$, $E_{7}$ with two conjugacy classes). In the last three columns the first three coefficients in the $\Theta$-series are listed with $a_{2}$ being a kissing number for a given lattice (except for Leech).}
\end{table}

\end{turnpage}

\end{document}